\documentclass[aps,preprint]{revtex4}
\usepackage{amssymb,epsf}
\usepackage{amsmath}
\usepackage{graphicx}

\begin{document}

\title{Nonlinear Charged Black holes in anti-de Sitter
Quasi-topological Gravity}
\author{ M. Ghanaatian $^{1}$ \footnote{%
email address: $m_{-}ghanaatian@pnu.ac.ir$} and A. Bazrafshan
$^{2}$} \affiliation{$^1$ Department of Physics, Payame Noor
University, Iran } \affiliation{$^2$ Department of Physics, Jahrom
University, 74137-66171 Jahrom, Iran}
\begin{abstract}
In this paper, we present the static charged solutions of quartic
quasitopological gravity in the presence of a nonlinear
electromagnetic field. Two  branches of these solutions present
black holes with one or two horizons or a naked singularity
depending on the charge and mass of the black hole. The entropy of
the charged black holes of fourth order quasitopological gravity
through the use of Wald formula is computed and the mass,
temperature and the charge of these black holes are found as well.
We show that black holes with spherical, flat and hyperbolical
horizon in quasitopological gravity are stable for any allowed
quasitopological parameters.
\end{abstract}

\maketitle

\section{Introduction}
One main reason for investigating the black holes of higher
dimensions  with a negative cosmological constant is finding the
correspondence between the gravitating fields in an anti-de Sitter
(AdS) space-time and conformal field theory living on the boundary
of the AdS space- time \cite{AdSCFT1}. The Ads/CFT correspondance
has become a key concept in the calculation of action and
thermodynamical quantities for every reference space-time
\cite{AdSCFT2,AdSCFT3,AdSCFT4,AdSCFT5}. This presumption has been
taken into account for asymptotically de Sitter space-times
\cite{AdSCFT6,AdSCFT7}.Most applications of the (A)dS/CFT
correspondence is for infinite boundaries, but sometimes it is
used for finite boundaries in order to obtain the conserved and
thermodynamic quantities
 \cite{AdSCFT8}.
One of the key issues, in recent years, has been has been the
thermodynamics of black holes in Ads space-time. If the AdS/CFT
correspondence be used, then, by studying the thermodynamics of
Ads black holes, we can certainly gain deep insights in the
characteristics and phase structures of strong 't Hooft coupling
CFTs by studying thermodynamics of AdS black holes. As regards the
topological characteristics, the asymptotically AdS black holes
are different from the asymptotically flat or dS black holes. One
of the main differences is that the black hole horizon topology
should be a round sphere S2 in four dimensional asymptotically
flat or dS spaces\cite{AdSCFT9}. It should be noted that,However,
the constant curvature of black holes horizon should be zero or
negative in asymptotically AdS spaces  . Many researchers have
investigated the thermodynamical properties of asymptotically
anti-de Sitter space-times with nonspherical horizon
\cite{AdSCFT4,AdSCFT10}. In Einstein gravity \cite{AdSCFT11}, the
black holes of Gauss-Bonnet gravity with hyperbolic horizon is
found stable \cite{AdSCFT12}. It seems that the Lovelock terms may
have no effect on the stability of topological black holes. In
Ref. \cite{AdSCFT13} the authors have shown that an asymptotically
flat uncharged black hole of third order Lovelock gravity may have
two horizons, which does not happen in lower order Lovelock
gravity. Also in Ref. \cite{AdSCFT14} the authors investigate the
effects of third order Lovelock term on the stability of a
spherical black hole of third order Lovelock gravity.

Some special features of higher curvature gravities have been
mentioned in Ref. \cite{AdSCFT15}. The asymptotically Ads black
hole solutions and their thermodynamic properties have also been
discussed in quartic quasitopological gravity in Ref.
\cite{AdSCFT11}. It should be noticed that the equations of motion
are only second order in derivatives ( for spherical symmetry) in
quartic quasitopological and the quartic topological action
\cite{AdSCFT15} yields nontrivial second-order field equations in
all space-time dimensionalities except 8 (beginning with the
five-dimensional). In Ref. \cite{AdSCFT16} the cubic
quasitopological is introduced. They investigate the holographic
properties of the  cubic Lagrangian as well, furthermore, they
prove that the generic perturbations around an AdS background for
such theory fulfill a second order equation. In Ref.
\cite{AdSCFT17} a Lagrangian out of a linear combination of cubic
invariants have been constructed which generically gives fourth
order field equations. But such field equations reduce to second
order when evaluated on spherically, hyperbolically or planar
symmetric spacetimes. They have also introduced a Lagrangian
containing quartic terms in the curvature with similar properties
to Ref. \cite{AdSCFT15} for dimensions greater that seven. Due of
special features of quartic quasi topological gravity in this
paper, we find the asymptotically Ads charged black hole solutions
and study the stability of these kinds of black holes. It is shown
that black holes of quasitopological gravity with arbitrary
curvature horizon are thermodynamically stable.

The outline of this paper is as follows: We introduce the action
of quartic quasitopological gravity in $(n+1)$ dimensions in the
presence of nonlinear electromagnetic field in Sec. (II). In
section (III), we obtain the general solutions of quasitopological
gravity. Section (IV) is devoted to calculate the entropy,
temprature and mass of charged black holes in quasitopological
gravity. We also investigate the stability of the charged black
holes with curved horizons in Sec. (V), and finally, we finish our
paper with some concluding remarks.

\section{CHARGED ACTION OF QUARTIC QUASI-TOPOLOGICAL GRAVITY IN $(n+1)$ DIMENSIONS } \label{3rdL}

The action of 4-th order quasitopological gravity in $(n+1)$
dimensions in the presence of the nonlinear electromagnetic field
can be formulated as follows:
\begin{equation}
I_{G}=\frac{1}{16\pi }\int d^{n+1}x\sqrt{-g}[-2\Lambda
+\mathcal{L}_{1}+\mu _{2}\mathcal{L}_{2}+\mu
_{3}\mathcal{X}_{3}+\mu _{4}\mathcal{X}_{4}+L(F)], \label{Act1}
\end{equation}%
where $\Lambda =-n(n-1)/2l^{2}$ is cosmological constant,
$\mathcal{L}_{1}={R}$ is the Einstein-Hilbert Lagrangian,
$\mathcal{L}_{2}=R_{abcd}{R}^{abcd}-4{R}_{ab}{R}^{ab}+{R}^{2}$ is
the Gauss-Bonnet Lagrangian, $\mathcal{X}_{3}$\ is the
curvature-cubed Lagrangian \cite{AdSCFT16}
\begin{eqnarray}
\mathcal{X}_{3} &=&R_{ab}^{cd}R_{cd}^{\,\,e\,\,\,f}R_{e\,\,f}^{\,\,a\,\,\,b}+%
\frac{1}{(2n-1)(n-3)}\left(
\frac{3(3n-5)}{8}R_{abcd}R^{abcd}R\right.  \notag
\\
&&-3(n-1)R_{abcd}R^{abc}{}_{e}R^{de}+3(n+1)R_{abcd}R^{ac}R^{bd}  \notag \\
&&\left. +\,6(n-1)R_{a}{}^{b}R_{b}{}^{c}R_{c}{}^{a}-\frac{3(3n-1)}{2}%
R_{a}^{\,\,b}R_{b}^{\,\,a}R+\frac{3(n+1)}{8}R^{3}\right)
\end{eqnarray}
$\mathcal{X}_{4}$ is the 4-th order Lagrangian of quasitopological
gravity \cite{AdSCFT15}:
\begin{eqnarray}
\mathcal{X}_{4}\hspace{-0.2cm} &=&\hspace{-0.2cm}c_{1}R_{abcd}R^{cdef}R_{%
\phantom{hg}{ef}%
}^{hg}R_{hg}{}^{ab}+c_{2}R_{abcd}R^{abcd}R_{ef}R^{ef}+c_{3}RR_{ab}R^{ac}R_{c}{}^{b}+c_{4}(R_{abcd}R^{abcd})^{2}
\notag \\
&&\hspace{-0.1cm}%
+c_{5}R_{ab}R^{ac}R_{cd}R^{db}+c_{6}RR_{abcd}R^{ac}R^{db}+c_{7}R_{abcd}R^{ac}R^{be}R_{%
\phantom{d}{e}}^{d}+c_{8}R_{abcd}R^{acef}R_{\phantom{b}{e}}^{b}R_{%
\phantom{d}{f}}^{d}  \notag \\
&&\hspace{-0.1cm}%
+c_{9}R_{abcd}R^{ac}R_{ef}R^{bedf}+c_{10}R^{4}+c_{11}R^{2}R_{abcd}R^{abcd}+c_{12}R^{2}R_{ab}R^{ab}
\notag \\
&&\hspace{-0.1cm}%
+c_{13}R_{abcd}R^{abef}R_{ef}{}_{g}^{c}R^{dg}+c_{14}R_{abcd}R^{aecf}R_{gehf}R^{gbhd},
\label{X4}
\end{eqnarray}
where
\begin{eqnarray*}
c_{1} &=&-\left( n-1\right) \left( {n}^{7}-3\,{n}^{6}-29\,{n}^{5}+170\,{n}%
^{4}-349\,{n}^{3}+348\,{n}^{2}-180\,n+36\right) , \\
c_{2} &=&-4\,\left( n-3\right) \left( 2\,{n}^{6}-20\,{n}^{5}+65\,{n}^{4}-81\,%
{n}^{3}+13\,{n}^{2}+45\,n-18\right) , \\
c_{3} &=&-64\,\left( n-1\right) \left( 3\,{n}^{2}-8\,n+3\right) \left( {n}%
^{2}-3\,n+3\right) , \\
c_{4} &=&-{(n}^{8}-6\,{n}^{7}+12\,{n}^{6}-22\,{n}^{5}+114\,{n}^{4}-345\,{n}%
^{3}+468\,{n}^{2}-270\,n+54), \\
c_{5} &=&16\,\left( n-1\right) \left( 10\,{n}^{4}-51\,{n}^{3}+93\,{n}%
^{2}-72\,n+18\right) , \\
c_{6} &=&--32\,\left( n-1\right) ^{2}\left( n-3\right) ^{2}\left( 3\,{n}%
^{2}-8\,n+3\right) , \\
c_{7} &=&64\,\left( n-2\right) \left( n-1\right) ^{2}\left( 4\,{n}^{3}-18\,{n%
}^{2}+27\,n-9\right) , \\
c_{8} &=&-96\,\left( n-1\right) \left( n-2\right) \left( 2\,{n}^{4}-7\,{n}%
^{3}+4\,{n}^{2}+6\,n-3\right) , \\
c_{9} &=&16\left( n-1\right) ^{3}\left( 2\,{n}^{4}-26\,{n}^{3}+93\,{n}%
^{2}-117\,n+36\right) , \\
c_{10} &=&{n}^{5}-31\,{n}^{4}+168\,{n}^{3}-360\,{n}^{2}+330\,n-90, \\
c_{11} &=&2\,(6\,{n}^{6}-67\,{n}^{5}+311\,{n}^{4}-742\,{n}^{3}+936\,{n}%
^{2}-576\,n+126), \\
c_{12} &=&8\,{(}7\,{n}^{5}-47\,{n}^{4}+121\,{n}^{3}-141\,{n}^{2}+63\,n-9), \\
c_{13} &=&16\,n\left( n-1\right) \left( n-2\right) \left(
n-3\right) \left(
3\,{n}^{2}-8\,n+3\right) , \\
c_{14} &=&8\,\left( n-1\right) \left( {n}^{7}-4\,{n}^{6}-15\,{n}^{5}+122\,{n}%
^{4}-287\,{n}^{3}+297\,{n}^{2}-126\,n+18\right) ,
\end{eqnarray*}
and $L(F)$ is the Lagrangian of power Maxwell invariant theory
\cite{AdSCFT18,AdSCFT19}
\begin{equation}
L(F)={(-F)}^{s}  \label{Lmat}
\end{equation}%
where $F_{\mu \nu }=\partial _{\mu }A_{\nu }-\partial _{\nu
}A_{\mu }$ is the electromagnetic tensor field and $A_{\mu }$ is
the vector potential. We should mention that in the limit $s=1$,
$L(F)$  reduces to the standard Maxwell Lagrangian.

\section{$(n + 1)$-DIMENSIONAL SOLUTIONS IN QUASI-TOPOLOGICAL GRAVITY }

The purpose of this section is to introduce the charged black hole
solutions of quasitopological gravity in the presence of nonlinear
Maxwell field. The action per unit volume can be written as,
\begin{equation}
I_{G}=\frac{{(n-1)}}{16\pi l^{2}}\int dtd\rho \lbrack {{{N(\rho
)\left[ \rho ^{n}(1+\psi +\hat{\mu}_{2}\psi ^{2}+\hat{\mu}_{3}\psi
^{3}+\hat{\mu}_{4}\psi
^{4})\right] ^{\prime }+\frac{2^{s}l^{2}\rho ^{(n-1)}h^{\prime 2s}}{%
(n-1)N(\rho )^{2s-1}}}}}],  \label{Act3}
\end{equation}%
where $\psi =l^{2}\rho^{-2}(k-f)$ and the dimensionless parameters $%
\hat{\mu}_{2}$, $\hat{\mu}_{3}$ and $\hat{\mu}_{4}$ are defined
as:
\begin{equation*}
\hat{\mu}_{2}\equiv \frac{(n-2)(n-3)}{l^{2}}\mu _{2},\text{ \ \ \ }\hat{\mu}%
_{3}\equiv \frac{(n-2)(n-5)(3n^{2}-9n+4)}{8(2n-1)l^{4}}\mu _{3},
\end{equation*}%
\begin{equation*}
\hat{\mu}_{4}\equiv {\frac{n\left( n-1\right) \left( n-2\right)
^{2}\left(
n-3\right) \left( n-7\right) ({{n}^{5}-15\,{n}^{4}+72\,{n}^{3}-156\,{n}%
^{2}+150\,n-42)}}{{l}^{6}}}\mu _{4},
\end{equation*}
In order to obtain the action (\ref{Act3}),the static spherically
symmetric metric and the vector potential as follows have been
used
\begin{equation}
ds^{2}=-N(\rho)^{2}f(\rho)dt^{2}+\frac{d\rho^{2}}{f(\rho)}+\rho^{2}d\Omega
^{2}, \label{metric}
\end{equation}
\begin{equation}
A_{\mu }=h(\rho )\delta _{\mu }^{0}
\end{equation}%
where
\[
d\Omega ^{2}=\left\{
\begin{array}{cc}
d\theta
_{1}^{2}+\sum\limits_{i=2}^{n-1}\prod\limits_{j=1}^{i-1}\sin
^{2}\theta _{j}d\theta _{i}^{2} & k=1 \\
d\theta _{1}^{2}+\sinh ^{2}\theta _{1}d\theta _{2}^{2}+\sinh
^{2}\theta _{1}\sum\limits_{i=3}^{n-1}\prod\limits_{j=2}^{i-1}\sin
^{2}\theta
_{j}d\theta _{i}^{2} & k=-1 \\
\sum\limits_{i=1}^{n-1}d\phi _{i}^{2} & k=0
\end{array}
\right\}\]
represents the line element of an $(n-1)$-dimensional
hypersurface with constant curvature $(n-1)(n-2)k$ and volume
$V_{n-1}$.
 Varying the action (\ref{Act3}) with respect to $f(\rho)$, we
obtain $N(\rho )=1$. Therefore, the variation with respect to
$h(\rho )$ gives
\begin{equation}
(2s-1)\rho h^{\prime \prime }+(n-1)h^{\prime }=0,  \label{eom2}
\end{equation}%
The solution of this differential equation is
\begin{equation}
h(\rho )=\left\{
\begin{array}{cc}

q\ln (\rho ), & s=n/2 \\
-q\rho ^{-(n-2s)/(2s-1)}, & 1/2<s<n/2%

\end{array}%
\right.  \label{hr}
\end{equation}%
here $q$ is an integration constant which is related to the charge
parameter. Since the potential should be finite at infinity for
$s\neq n/2 $, the interval $1/2<s<n/2$ is chosen. One can find the
electric charge according to the Gauss theorem as:
\begin{equation}
Q=\frac{1}{4\pi}\int_{r\rightarrow\infty}F_{t\rho}\sqrt{-g}d^{n-1}x=\frac{2^{s}s{(n-2s)^{2s-1}}V_{n-1}q^{2s-1}}{8\pi
{(2s-1)}^{2s-1}}
\end{equation}
 varying the action (\ref{Act3}) with respect to
$N(\rho)$ yields
\begin{equation}
\psi ^{4}+\frac{\hat{\mu}_{3}}{\hat{\mu}_{4}}\psi ^{3}+\frac{\hat{\mu}_{2}}{%
\hat{\mu}_{4}}\psi ^{2}+\frac{1}{\hat{\mu}_{4}}\psi +\frac{1}{\hat{\mu}_{4}}%
\kappa =0,  \label{Eq4}
\end{equation}
where
\begin{equation}
\kappa =1-\frac{m}{\rho ^{n}}+\frac{2^{s}l^{2}q^{2s}(n-2s)^{2s-1}}{%
(n-1)(2s-1)^{(2s-2)}\rho ^{2s(n-1)/(2s-1)}},
\end{equation}
and $m$ is an integration constant which can be evaluated as the
geometrical mass of black hole solutions in terms of the horizon
radius
\begin{equation}
m=\left( 1+k\frac{l^{2}}{\rho_{+}^{2}}+\hat{\mu }_{2}{k}^{2}\frac{%
l^{4}}{\rho_{+}^{4}}+\hat{\mu }_{3}{k}^{3}\frac{l^{6}}{\rho_{+}^{6}}+\hat{\mu }_{4}%
{k}^{4}\frac{l^{8}}{\rho_{+}^{8}}+\frac{%
2^{s}l^{2}q^{2s}(n-2s)^{2s-1}(\rho_{+})^{2s(1-n)/(2s-1)}}{%
(n-1)(2s-1)^{(2s-2)}} \right) {\rho_{+}^{n}} \label{mh}
\end{equation}
In order to find the black hole solutions, we choose two solutions
of $f(\rho)$ as
\begin{equation}
f_{1}(\rho)=k+\frac{\rho^{2}}{l^{2}}\left( \frac{\hat{\mu}_{3}}{4\hat{\mu}_{4}}+\frac{1%
}{2}R-\frac{1}{2}E\right) .  \label{Fr4}
\end{equation}
\begin{equation}
f_{2}(\rho)=k+\frac{\rho^{2}}{l^{2}}\left( \frac{\hat{\mu}_{3}}{4\hat{\mu}_{4}}-\frac{1%
}{2}R+\frac{1}{2}K\right) .  \label{Fr4}
\end{equation}
where
\begin{eqnarray}
R &=&\left( \frac{{\hat{\mu}_{3}}^{2}}{4{\hat{\mu}_{4}}^{2}}-\frac{\hat{\mu}%
_{2}}{\hat{\mu}_{4}}+y_{1}\right) ^{1/2},
\label{RR} \\
E &=&\left( \frac{3{\hat{\mu}_{3}}^{2}}{4{\hat{\mu}_{4}}^{2}}-\frac{2\hat{\mu%
}_{2}}{\hat{\mu}_{4}}-R^{2}-\frac{1}{4R}\left[ \frac{4\hat{\mu}_{2}\hat{\mu}%
_{3}}{{\hat{\mu}_{4}}^{2}}-\frac{8}{\hat{\mu}_{4}}-\frac{{\hat{\mu}_{3}}^{3}%
}{{\hat{\mu}_{4}}^{3}}\right] \right) ^{1/2},  \label{EE} \\
K &=&\left( \frac{3{\hat{\mu}_{3}}^{2}}{4{\hat{\mu}_{4}}^{2}}-\frac{2\hat{\mu%
}_{2}}{\hat{\mu}_{4}}-R^{2}+\frac{1}{4R}\left[ \frac{4\hat{\mu}_{2}\hat{\mu}%
_{3}}{{\hat{\mu}_{4}}^{2}}-\frac{8}{\hat{\mu}_{4}}-\frac{{\hat{\mu}_{3}}^{3}%
}{{\hat{\mu}_{4}}^{3}}\right] \right) ^{1/2}  \label{KK}\\
\Delta &=&\frac{H^{3}}{27}+\frac{D^{2}}{4},\text{ \ \ \ \ \ \ }H={\frac{3%
\hat{\mu}_{3}-{\hat{\mu}_{2}}^{2}}{3{\hat{\mu}_{4}}^{2}}}-\,{\frac{4\kappa }{%
\hat{\mu}_{4}},} \\
D &=&{\frac{2}{27}}\,{\frac{{\hat{\mu}_{2}}^{3}}{{\hat{\mu}_{4}}^{3}}}-\frac{%
1}{3}\,\left( {\frac{\hat{\mu}_{3}}{{\hat{\mu}_{4}}^{2}}}+8\,{\frac{\kappa }{%
\hat{\mu}_{4}}}\right) \frac{\hat{\mu}_{2}}{\hat{\mu}_{4}}+{\frac{{\hat{\mu}%
_{3}}^{2}\kappa
}{{\hat{\mu}_{4}}^{3}}}+\frac{1}{{\hat{\mu}_{4}}^{2}}.
\end{eqnarray}
and $y_{1}$ is the real root of following equation:
\begin{equation}
{y}^{3}-{\frac {\mu_{{2}}{y}^{2}}{\mu_{{4}}}}+ \left( {\frac
{\mu_{{3} }}{{\mu_{{4}}}^{2}}}-4\,{\frac {\kappa}{\mu_{{4}}}}
\right) y-{\frac {
{\mu_{{3}}}^{2}\kappa}{{\mu_{{4}}}^{3}}}+\,{\frac
{4\mu_{{2}}\kappa}{{ \mu_{{4}}}^{2}}}-\frac{1}{{\mu_{{4}}}^{2}}=0
\end{equation}
The metric function $f(\rho)$ for the uncharged solution $(q=0)$
is real in the whole range $0\leq \rho <\infty$. But for the
charged solution, the spacetime should be restricted to the region
$\rho\geq r_{0}$, we introduce a new radial coordinate r as
\begin{equation}
r=\sqrt{{\rho}^{2}-{r_{0}}^{2}}\Rightarrow
d{\rho}^{2}=\frac{r^{2}}{r^{2}+r_{0}^{2}}dr^{2}
\end{equation}
 where $r_{0}$ is the largest real root of
 $\Delta_{0}=\Delta(\kappa=\kappa_{0})$,
 $R_{0}=R(\kappa=\kappa_{0})$,
 $E_{0}=E(\kappa=\kappa_{0})$ and
 $K_{0}=K(\kappa=\kappa_{0})$, and $\kappa_{0}$ is
\begin{equation}
\kappa_{0}
=1-\frac{ml^{2}}{r_{0}^{n}}+\frac{2l^{2}(n-2)}{(n-1)}\frac{q^{2}}{r_{0}^{2(n-1)}}
\label{kap}
\end{equation}
So the metric of an ($n+1$)-dimensional static spherically
symmetric spacetime changes to:
\begin{equation}
ds^{2}=-f(r)dt^{2}+\frac{r^{2}dr^{2}}{f(r)(r^{2}+r_{0}^{2})}+(r^{2}+r_{0}^{2})d\Omega
^{2} \label{metric}
\end{equation}
and the functions $h(r)$ and $\kappa $ changes to
\begin{eqnarray}
h(r) &=&-q(r^{2}+r_{0}^{2})^{(2s-n)/(4s-2)}  \label{hr2} \\
\kappa &=&1-\frac{m}{(r^{2}+r_{0}^{2})^{n/2}}+\frac{%
2^{s}l^{2}q^{2s}(n-2s)^{2s-1}}{%
(n-1)(2s-1)^{(2s-2)}(r^{2}+r_{0}^{2})^{s(n-1)/(2s-1)}},
\end{eqnarray}%

\section{THE THERMODYNAMICS QUANTITIES OF THE BLACK HOLES}

In order to use Wald's formula \cite{AdSCFT20}, one can evaluate
the entropy per unit volume $V_{n-1}$ \cite{AdSCFT15}:
\begin{equation}
\mathcal{S}=\frac{{\eta_{+}^{n-1}}}{4}\left(
1+2\,k\hat{\mu}_{2}{\frac{\left(
n-1\right) {l}^{2}}{\left( n-3\right) \eta_{+}^{2}}}+3k^{2}\hat{\mu}_{3}{\frac{%
\left( n-1\right) {\ l}^{4}}{\left( n-5\right) \eta_{+}^{4}}}+4{k}^{3}\hat{\mu}%
_{4}{\frac{\left( n-1\right) {l}^{6}}{\left( n-7\right)
\eta_{+}^{6}}}\right) \label{SS}
\end{equation}
where $\eta_{+}=\sqrt{r_{+}^{2}+r_{0}^{2}}$. This entropy reduces
to the area law of entropy for
$\hat{\mu}_{2}=\hat{\mu}_{3}=\hat{\mu}_{4}=0$ (Einstein gravity).
furthermore the temperature of the event horizon by the standard
method of analytic continuation of the metric can be calculated as
\begin{eqnarray}
T&=&\frac{f^{\prime }(r_{+})}{4\pi
}\sqrt{1+\frac{r_{0}^{2}}{r_{+}^{2}}} \notag
\\
&&={\frac{n\hat{\mu}%
_{0}\eta_{+}^{8}+\left( n-2\right) k{l}^{2}\eta_{+}^{6}+\left( n-4\right) {k}^{2}%
\hat{\mu}_{2}l^{4}\eta_{+}^{4}+\left( n-6\right) k\hat{\mu}_{3}{l}%
^{6}\eta_{+}^{2}+\left( n-8\right)
{k}^{2}\hat{\mu}_{4}{l}^{8}}{\left(
\,\eta_{+}^{6}+2k\hat{\mu}_{2}{l}^{2}\eta_{+}^{4}\,+3k^{2}\hat{\mu}_{3}{l}%
^{4}\eta_{+}^{2}\,+4\hat{\mu}_{4}k^{3}{l}^{6}\right) 4\pi
\,{l}^{2}\eta_{+}}} \notag
\\
&&-\frac{q^{2s}2^{s}{(n-2s)}^{2s}(\eta_{+})^{{%
2s(1-n)}/{(2s-1)}}}{4\pi l^{2} \left(
4\,\hat{\mu}_{4}{k}{\eta_{+}}^{-6}{l}^{6}+3\,{\eta_{+}}^{-4}\hat{\mu}_{3}{k}^{2}{l}^{
4}+2\,{\eta_{+}}^{-2}\hat{\mu}_{2}k{l}^{2}+1 \right)
 (n-1)(2s-1)^{2s-1}}\eta_{+}\label{TT}
\end{eqnarray}

The electric potential $\Phi $, which is measured at infinity with
respect to the horizon, is introduced by \cite{AdSCFT21,AdSCFT22}
\begin{equation}
\Phi =A_{\mu }\xi ^{\mu }\mid _{r\rightarrow \infty }-A_{\mu }\xi
^{\mu }\mid _{r=r_{+}}  \label{Pot}
\end{equation}%
where $\xi ^{\mu }$ is the null generator of the horizon, $A_{\mu
}$ is the vector potential. Therefore the following can be
calculated
\begin{equation}
\Phi =\frac{q}{(r_{+}^{2}+r_{0}^{2})^{(n-2s)/2(2s-1)}} \label{Ph0}
\end{equation}%

The ADM mass of black hole can be obtained by using the behavior
of the metric at large $r$. One can show that the mass of the
black hole per unit volume, $V_{n-1}$, is
\begin{equation}
M=\frac{(n-1)}{16\pi}m \label{bent}
\end{equation}
meanwhile the first law of thermodynamics, $dM=TdS+\Phi dQ$, is
satisfied. This is due to the fact that the intensive quantities
\begin{equation}
T=\left( \frac{\partial M}{\partial S}\right) _{Q},\,\ \ \ \ \Phi
=\left( \frac{\partial M}{\partial Q}\right) _{S}  \label{Inten}
\end{equation}%
are in agreement with those given in Eqs. (\ref{TT}) and
(\ref{Ph0}).

\begin{figure}[ht]
\centering {\includegraphics[width=7cm]{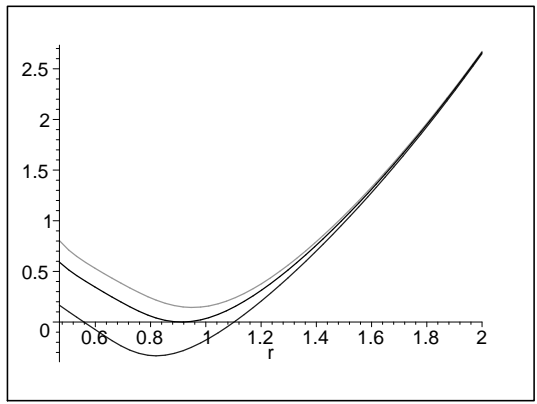} }
\caption{$f_{1}(r)$ vs. $r$ for $k=-1$, $l=1$,$n=4$,
$\hat{\mu}_{0}=1$, $\hat{\mu}_{2}=.1$, $\hat{\mu}_{3}=.2$,
$\hat{\mu}_{4}=.01$, $m=.5$ and $q>q_{\mathrm{ext}}$,
$q=q_{\mathrm{ext}}$ and $q<q_{\mathrm{ext}}$ from up to down,
respectively.} \label{Fr7a}
\end{figure}

\begin{figure}[ht]
\centering {\includegraphics[width=7cm]{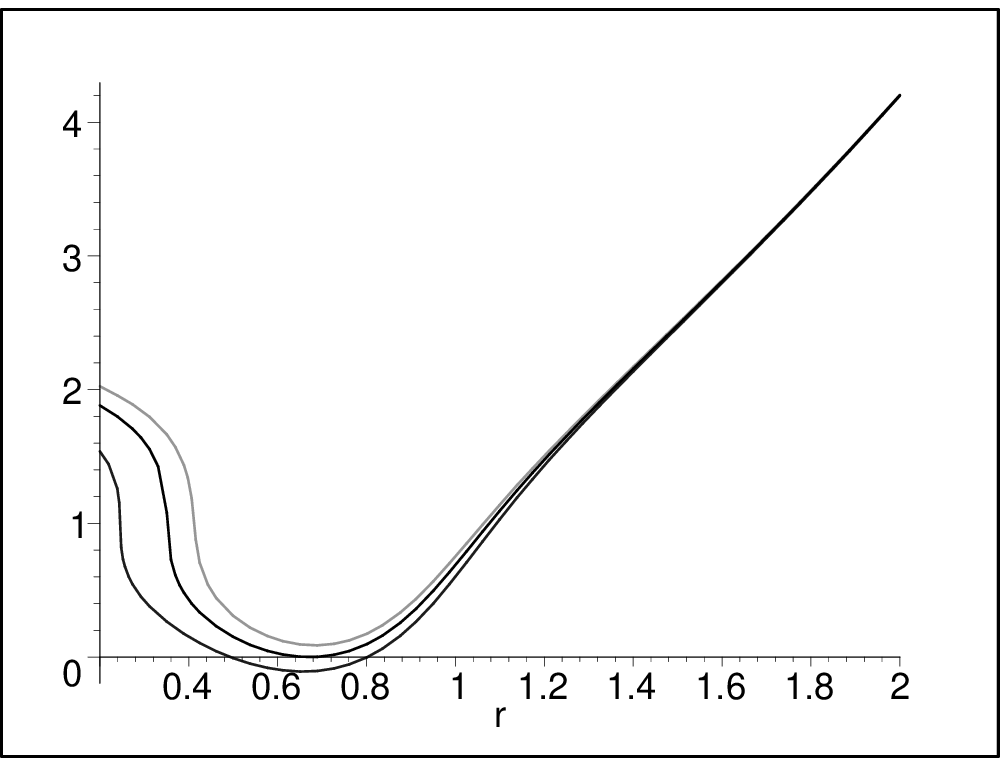} }
\caption{$f_{1}(r)$ vs. $r$ for $k=1$, $l=1$,$n=4$,
$\hat{\mu}_{0}=1$, $\hat{\mu}_{2}=-.01$, $\hat{\mu}_{3}=.2$,
$\hat{\mu}_{4}=.001$, $m=1.5$ and $q>q_{\mathrm{ext}}$,
$q=q_{\mathrm{ext}}$ and $q<q_{\mathrm{ext}}$ from up to down,
respectively.} \label{Fr7b}
\end{figure}

Note that all the thermodynamic quantities obtained in this
section reduce to those of Einstein gravity given in
\cite{AdSCFT11} for $\hat{\mu}_{2}=\hat{\mu}_{3}=\hat{\mu}_{4}=0$.
 Figures \ref{Fr7a} and Figures \ref{Fr7b}
show the metric function $f_{1}(r)$ for different values of charge
parameters with $k=-1$ and $k=+1$ respectively. The charge of the
extreme black hole may be obtained by using Eq. (\ref{mh}) and
computing $q_{\mathrm{ext}}=q(r_{\mathrm{ext}})$.
Then, our solution presents a black hole with inner and outer horizons provided $q>q_{\mathrm{%
ext}}$, an extreme black hole if $q=q_{\mathrm{ext}}$, and a naked
singularity for $q<q_{\mathrm{ext}}$. Figure \ref{Fr7c} shows the
metric function $f_{2}(r)$ for different values of charge
parameters with $k=-1$. Therefore, the charge of the extreme black
hole may be obtained by using Eq. (\ref{mh}) and evaluating
$q_{\mathrm{ext}}=q(r_{\mathrm{ext}})$.

\section{STABILITY IN THE CANONICAL ENSEMBLE}

\begin{figure}[ht]
\centering {\includegraphics[width=7cm]{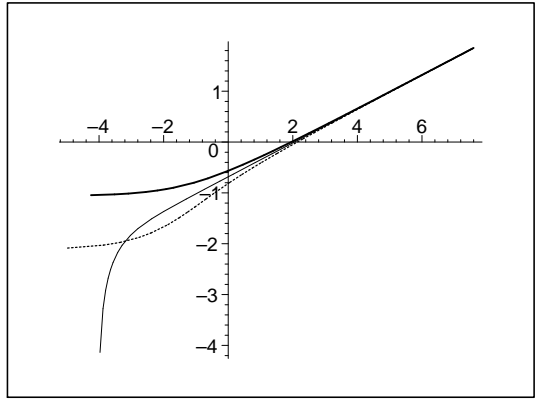} }
\caption{$logS$ vs. $logT$ for $n=4$, $q=.1$, $l=1$,
$\hat{\mu}_{2}=.2$, $\hat{\mu}_{3}=.1$, $\hat{\mu}_{4}=.007$,
$k=1$(dashed line), $k=0$(solid line) and $k=-1$(bold line).}
\label{Fr7d}
\end{figure}

\begin{figure}[ht]
\centering {\includegraphics[width=7cm]{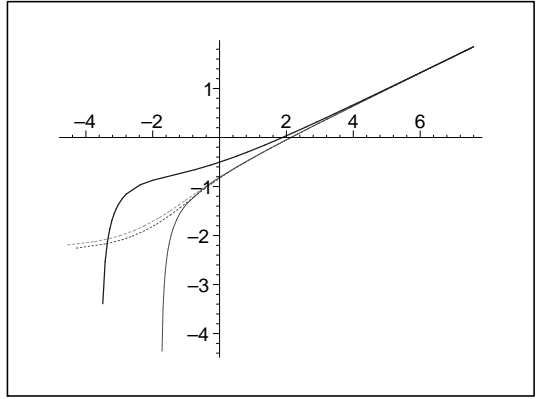} }
\caption{$logS$ vs. $logT$ for $n=4$, $q=.1$, $l=1$,
$\hat{\mu}_{2}=.2$, $\hat{\mu}_{3}=.1$, $\hat{\mu}_{4}=.06$,
Quartic-quasi (dashed line), Cubic-quasi (dotted line),
Gauss-Bonnet(solid line) and Einsteinian (bold line).}
\label{Fr7e}
\end{figure}

The local stability analysis in any ensemble can be modified by
finding the determinant of the Hessian matrix
$|{\partial^{2}S}/{\partial X_{i}\partial X_{j}}|$, where
$X_{i}$'s are the thermodynamic variables of the system
\cite{AdSCFT21}. The number of the thermodynamic variables depends
on the ensemble which is used. Since the charge is a fixed
parameter in the canonical ensemble, therefore, the positivity of
the thermal capacity  $C_{Q}$ is a sufficient factor to have the
local stability. The thermal capacity $C_{Q}$ at constant charge
is
\begin{equation}
C_{Q}=T\frac{\partial S}{\partial T}
\end{equation}

as a result, the slope of $logS$ versus $logT$ shows the
thermodynamic stability of the black holes. For $k=0$ the
temperature and the entropy do not depend on quasitopological
coefficients. Figures. \ref{Fr7d} and \ref{Fr7e} show that the
black holes with flat horizons are stable. We will discuss the
stability of hyperbolic and spherical black holes in the rest of
this paper.

\begin{figure}[ht]
\centering {\includegraphics[width=7cm]{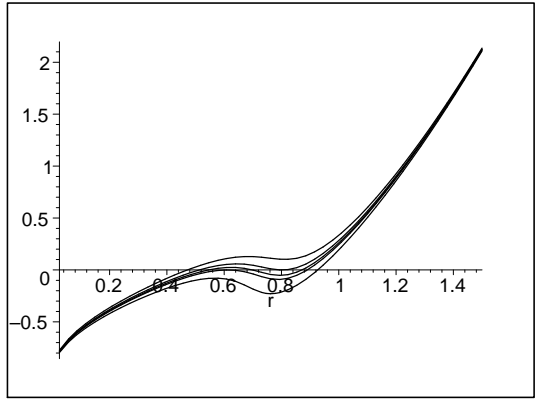} }
\caption{$f_{2}(r)$ vs. $r$ for $k=-1$, $l=1$,$n=4$,
$\hat{\mu}_{0}=1$, $\hat{\mu}_{2}=.2$, $\hat{\mu}_{3}=-.1$,
$\hat{\mu}_{4}=-.06$, $m=.3$ and $q>q_{1\mathrm{ext}}$,
$q=q_{1\mathrm{ext}}$, $q_{2\mathrm{ext}}<q<q_{1\mathrm{ext}}$,
$q=q_{2\mathrm{ext}}$ and $q<q_{2\mathrm{ext}}$ from up to down,
respectively.} \label{Fr7c}
\end{figure}

It should be mentioned that the action of quartic quasitopological
gravity contributes to the equations of motion in any dimensions
except 8. Figure \ref{Fr7c} shows the metric function $f_{2}(r)$
for different values of charge parameters with $k=-1$. In the case
of $k=-1$, it can be seen that there are two values for the charge
of the extreme black hole $q_{\mathrm{1ext}}=q(r_{\mathrm{1ext}})$
and $q_{\mathrm{2ext}}=q(r_{\mathrm{2ext}})$. For black holes with
$q<q_{\mathrm{2ext}}$, the charge of the extreme black hole is
$q_{\mathrm{1ext}}$, while for black holes with
$q>q_{\mathrm{2ext}}$, the charge of the extreme black hole is
$q_{\mathrm{2ext}}$. Where $r_{\mathrm{1ext}}$ and
$r_{\mathrm{2ext}}$  are real root(s) for the equation $T=0$.

In the case of $k=-1$ the critical charge $q_{crit}$, can be
calculated numerically, if we choose $q<q_{crit}$ the function
$f_{2}(r)$ to become imaginary. For hyperbolic black holes, the
$logS$ versus $logT$ is shown in Fig. \ref{Fr7d}, in which the
slope of the $logS$ versus $\log T$ is always positive, and
therefore these black holes are thermodynamically stable.


The solution for $k=1$ presents a black hole with inner and outer horizons provided $q>q_{\mathrm{%
ext}}$, an extreme black hole if $q=q_{\mathrm{ext}}$, and a naked
singularity for $q<q_{\mathrm{ext}}$.

It is known that the negative slope in a temperature-entropy plot
produces a situation in which the black hole will be
thermodynamically unstable. Considering AdS black holes , we learn
from Fig.\ref{Fr7d} that no phase transition takes place for
spherical black holes because the slope is positive in all aspect.
The most important result of this paper is to clarify this fact
that quartic quasitopological term has no effect on the stability
of the black holes (Fig. \ref{Fr7e}).

\section{Concluding Remarks}

We calculated the entropy and the temperature of quasitopological
black holes and found that the entropy of fourth gravity reduces
to the area law of the entropy for
$\hat{\mu_{2}}=\hat{\mu_{3}}=\hat{\mu_{4}}=0$. Using Gauss
theorem, we also calculate the charge of the black hole. In
addition, it is shown that the first law of thermodynamics is
satisfied. For the case $k=1$ and $k=-1$,
we obtained the solution that presents a black hole with inner and outer horizons provided $q>q_{\mathrm{%
ext}}$, an extreme black hole if $q=q_{\mathrm{ext}}$, and a naked
singularity for $q<q_{\mathrm{ext}}$ . At the same time,  for
quasitopological black holes with hyperbolic horizon is obtained
two values for the charge of the extreme black hole
$q_{\mathrm{1ext}}=q(r_{\mathrm{1ext}})$ and
$q_{\mathrm{2ext}}=q(r_{\mathrm{2ext}})$. For black holes with
$q<q_{\mathrm{2ext}}$, the charge of the extreme black hole is
$q_{\mathrm{1ext}}$, while for black holes with
$q>q_{\mathrm{2ext}}$, the charge of the extreme black hole is
$q_{\mathrm{2ext}}$. In Einstein and Gauss-Bonnet gravities, the
topological black holes with hyperbolic horizon  are stable
\cite{AdSCFT11,AdSCFT12}. The Lovelock terms, for black holes with
flat horizons, do not change the stability phase structure
\cite{AdSCFT23}.In this paper, we can show that the black holes in
quartic quasitopological gravity are thermodynamically stable.

\acknowledgments This work has been supported by  Payame Noor and
Jahrom universities.

\end{document}